\documentclass[twocolumn,aps,superscriptaddress,longbibliography]{revtex4-2}
\usepackage{bm}
\usepackage{amsmath}
\usepackage{amssymb}
\usepackage{times}
\usepackage[usenames,dvipsnames]{xcolor}
\usepackage{graphicx}
\usepackage{dcolumn}
\usepackage{simplewick}
\usepackage{hyperref}

\hypersetup{
  colorlinks=false,
  colorlinks=true,
  citecolor=blue,
  linkcolor=blue,
  urlcolor=blue}

\DeclareUnicodeCharacter{2212}{-}

\begin{document}

\title{Emergence of half-metallic ferromagnetism and valley polarization 
in transition metal substituted WSTe monolayer}

\author{Shivani Kumawat}
\affiliation{Department of Physics, Indian Institute of Technology,
             Hauz Khas, New Delhi 110016, India}

\author{Chandan Kumar Vishwakarma}
\affiliation{Materials Department, University of California, Santa Barbara, California 93106-5050, USA}

\author{Mohd Zeeshan}
\affiliation{Department of Physics, Indian Institute of Technology,
             Hauz Khas, New Delhi 110016, India}

\author{Indranil Mal}
\affiliation{Department of Physics, Indian Institute of Technology,
             Hauz Khas, New Delhi 110016, India}

\author{Sunil Kumar}
\email{kumarsunil@physics.iitd.ac.in}
\affiliation{Department of Physics, Indian Institute of Technology,
             Hauz Khas, New Delhi 110016, India}

\author{B. K. Mani}
\email{bkmani@physics.iitd.ac.in}
\affiliation{Department of Physics, Indian Institute of Technology,
             Hauz Khas, New Delhi 110016, India}

\begin{abstract}

Two-dimensional (2D) Janus materials hold a great importance in spintronic 
and valleytronic applications due to their unique lattice structures and 
emergent properties. 
They intrinsically exhibit both an in-plane inversion and out-of-plane 
mirror symmetry breakings, which offer a new degree of freedom to electrons 
in the material. One of the main limitations in the multifunctional 
applications of these materials 
is, however, that, they are usually non-magnetic in nature. 
Here, using first-principles calculations, we propose to induce magnetic 
degree of freedom in non-magnetic WSTe via doping with transition 
metal (TM) elements -- Fe, Mn and Co. Further, we comprehensively probe 
the electronic, spintronic and valleytronic properties in these systems.
Our simulations predict intrinsic Rashba and Zeeman-type spin splittings 
in pristine WSTe. The obtained Rashba parameter is $\sim$ 422 meV\AA\; along the 
$\Gamma - K$ direction. Our study shows a strong dependence on uniaxial 
and biaxial strains where we observe an enhancement of $\sim$ 2.1\% 
with 3\% biaxial compressive strain. The electronic structure of TM-substituted WSTe 
reveals half-metallic nature for 6.25 and 18.75\% of Fe, 25\% of Mn, and 
18.75 and 25\% of Co structures, which leads to 100\% spin polarization. 
The structural asymmetry, strong SOC, and broken inversion symmetry are 
found to lead to valley polarization in TM-WSTe systems. The obtained values
of valley polarization 65, 54.4 and 46.3 meV for 6.25\% of Fe, Mn and Co, 
respectively, are consistent with the literature data for other Janus 
materials. Further, our calculations show a strain dependent tunability 
of valley polarization, where we find an increasing (decreasing) trend 
with uniaxial and biaxial tensile (compressive) strains. We observed a 
maximum enhancement of $\sim$ 1.72\% for 6.25\% of Fe on 
application of 3\% biaxial tensile strain.
The ability to control valley degrees of freedom, along with spin and 
charge, could open a new prospect for next-generation spintronic 
and valleytronic devices. 
        
\end{abstract}

\date{\today}

\maketitle

\section{Introduction}

Analogous to the charge in electronics, spin degree of freedom (SDF) 
plays a crucial role in the field of spintronics \cite{ahn20202d}.
Owing to their much faster speed, ultra-low heat dissipation and non-voltality, 
spintronic devices are more suitable in comparison to their electronic 
counterparts \cite{avsar2020colloquium}. One of the key features of a spintronic material is the 
spin polarization, which is attributed to the differences in the population 
of spin-up and spin-down electrons. 
2D materials have been demonstrated show significant advantages for 
spintronic applications \cite{liu2020spintronics, PhysRevB.105.075414}. One of the primary 
benefits with 2D materials is their ability to exhibit intrinsic magnetism 
and half-metallicity at nanoscale. 
Experimentally, the room temperature ferromagnetism has been observed 
in transition metals doped 2D materials -- Fe-doped MoS$_2$ \cite{fu2020enabling}, 
V-doped WSe$_2$ \cite{yun2020ferromagnetic}, V-doped MoTe$_2$ \cite{coelho2019room}, 
Fe-doped ZrS$_2$ \cite{muhammad2018room}, TM (V, Cr, Mn, Fe, Co, Ni, and Cu)-
doped black phosphorus \cite{jiang2018room}.

In recent years, a new quantum degree of freedom, the valley degree of 
freedom (VDF), has emerged as a promising basis for device 
applications \cite{luo2024valleytronics, PhysRevLett.99.236809, zeng2012valley, PhysRevB.92.121403}.
The energy levels of electrons in valence and conduction bands of a 
2D material exhibit multiply degenerate extrema, known as 
valleys \cite{soni2022valley,vitale2018valleytronics}. The VDF based functional 
properties provide a roadmap for investigating the phenomena like optical 
circular dichroism, valley Hall effect and spin-valley 
locking \cite{khanikaev2016experimental, Mak, saito2016} in the materials.
One of key prerequisites in the practical realization of VDF based 
applications is, however, the precise production and control of valley 
polarization. The valley polarization can be induced in 2D materials 
by lifting the energy degeneracy at $K$ and $K^\prime$ valleys, and for this
magnetic doping is suggested to be one of the efficient
mechanisms \cite{cheng2014valley}. In terms of experiments, the first demonstration 
of valley related physics in a 2D material was presented in the case 
of graphene \cite{PhysRevLett.99.236809}. Another key experiment, 
Ref. \cite{zeng2012valley,mak2012control}, demonstrated inducing and controlling 
valley polarization in MoS$_2$ monolayer using circularly polarized light. 
Further, in a recent experiment, Sahoo {\em et al.} reported a high 
degree of valley polarization at room-temperature in Vanadium-doped 
MoS$_2$ \cite{PhysRevMaterials.6.085202}. In terms of simulations, 
Refs. \cite{acs.jpclett.8b01625} (MoSSe) and \cite{ZHAO2019172} 
(WSSe) have shown that the valley polarization can be induced in Janus 
2D materials via transition metal dopings. Considering the previous 
experimental and theoretical studies in literature, it can thus be surmised 
that the TM-doping in non-magnetic 2D materials could provide a mechanism 
to induce spintronics and valleytronics properties.

The present work aims to explore and propose a material which could offer 
both spintronics and valleytronic properties. For this, we have identified 
a new class of 2D materials, Janus-transition-metal-dichalcogenides 
(Janus-TMDCs), WSTe, and its doping with transition metal elements.
The reason for choosing this material for our study is, a research gap in 
terms of the understanding the effect of TM-doping on the properties of WSTe. 
It is to be mentioned that, its parent compound 
WTe$_2$ is reported to show a giant valley polarization, which is large enough 
to realize anomalous valley Hall effect, when doped with Co \cite{zhao2020enhanced}. 
It should be noted that, due to their unique lattice structure and the 
absence of out-of-plane mirror symmetry, Janus-TMDCs and based heterostructures 
have gained a tremendous attention for spintronic and valleytronics 
applications \cite{lv2022spin,wang2019janus,liu2022janus}.

More precisely, in this work, with the help of the state-of-the-art 
first-principles calculations, we aim to probe electronic, spintronic and 
valleytronic properties in transition metals (Fe, Mn and Co)-substituted 
Janus WSTe monolayer. To probe it in a comprehensive way, we aim to address
the following properties:
i) The nature of electronic structure of pristine and TM-substituted WSTe (TM-WSTe) monolayer.
ii) Rashba and Zeeman spin splittings in WSTe and their tunability using strain.
iii) The impact of TM-substitution in terms of induced magnetic properties in WSTe.
iv) Mechanism behind the advent of magnetic degrees of freedom in TM-WSTe
and spin polarization.
v) Probing the valley polarization and its tunability using uniaxial and 
biaxial strains.

The paper is organized into four sections. In Sec. II, we provide a brief 
description of the computational methods and parameters used in our
calculations. In Sec. III, we present and analyze our results on 
electronic structure, magnetic properties, Rashba and Zeeman spin 
splittings, and valley polarization in TM-substituted WSTe.
The summary of our results is presented in the last section of 
the paper.

\section{Computational Methodology}

To investigate the properties of WSTe and TM-WSTe from first-principles, we 
performed the density functional theory (DFT) based calculations using 
{\it Vienna Ab initio Simulation Package}
(VASP) package \cite{PhysRevB.59.1758, PhysRevB.49.14251, PhysRevB.54.11169}.
We used generalized gradient approximation (GGA) based Perdew-Burke-Ernzerhof 
(PBE) pseudopotential \cite{PhysRevLett.77.3865} to incorporate the 
exchange-correlations among electrons. The plane wave basis with an energy 
cutoff of 500 eV and the projector augmented wave \cite{PhysRevB.50.17953} to represent 
atomic cores were employed in all the calculations. For all self-consistent-field 
calculations we used energy and force convergence criteria of 
$10^{-6}$ eV and $10^{-4}$ eV/\AA, respectively. The $\Gamma$-centered 
$k$-mesh of 13$\times$13$\times$1 was used in all calculations. The computed
properties were, however, tested for convergence using higher $k$-mesh 
values. The effect of strongly correlated $d$-electrons of Fe/Mn/Co in TM-WSTe was 
incorporated using the rotationally invariant DFT+U approach 
of Dudarev {\it et al.} \cite{PhysRevB.57.1505}. The Hubbard U parameters were 
calculated self-consistently using density functional perturbation 
theory (DFPT) employing the cococcioni's approach \cite{PhysRevB.71.035105}.
Our computed values 4.4, 4.6, and 5.3 eV for Fe, Mn, and Co, respectively, 
are consistent with the values 4.6, 4.0, and 5.0 eV reported in the 
literature \cite{calderon2015aflow}. The small difference could be 
attributed to the difference in the pseudopotentials used and methods 
employed in the two calculations.

Further, to probe the Rashba and Zeeman-spin splittings, and valley 
polarization, we incorporated the relativistic effects using spin-orbit 
coupling (SOC) in our calculations. To avoid any artificial interaction 
between layers, a 12 {\AA} thick vacuum was introduced along the 
$z$-direction in WSTe and TM-WSTe structures. To incorporate the 
substitution of transition metals at W site, a 4$\times$4$\times$1 
supercell of WSTe monolayer was used. Considering the data available 
in literature and increasing computational complexity with smaller 
concentrations, we chose TM concentrations as 6.25, 12.50, 18.75, and 25\% 
in our study. 

\begin{figure}
\includegraphics[width=1
\columnwidth,angle=0,clip=true]{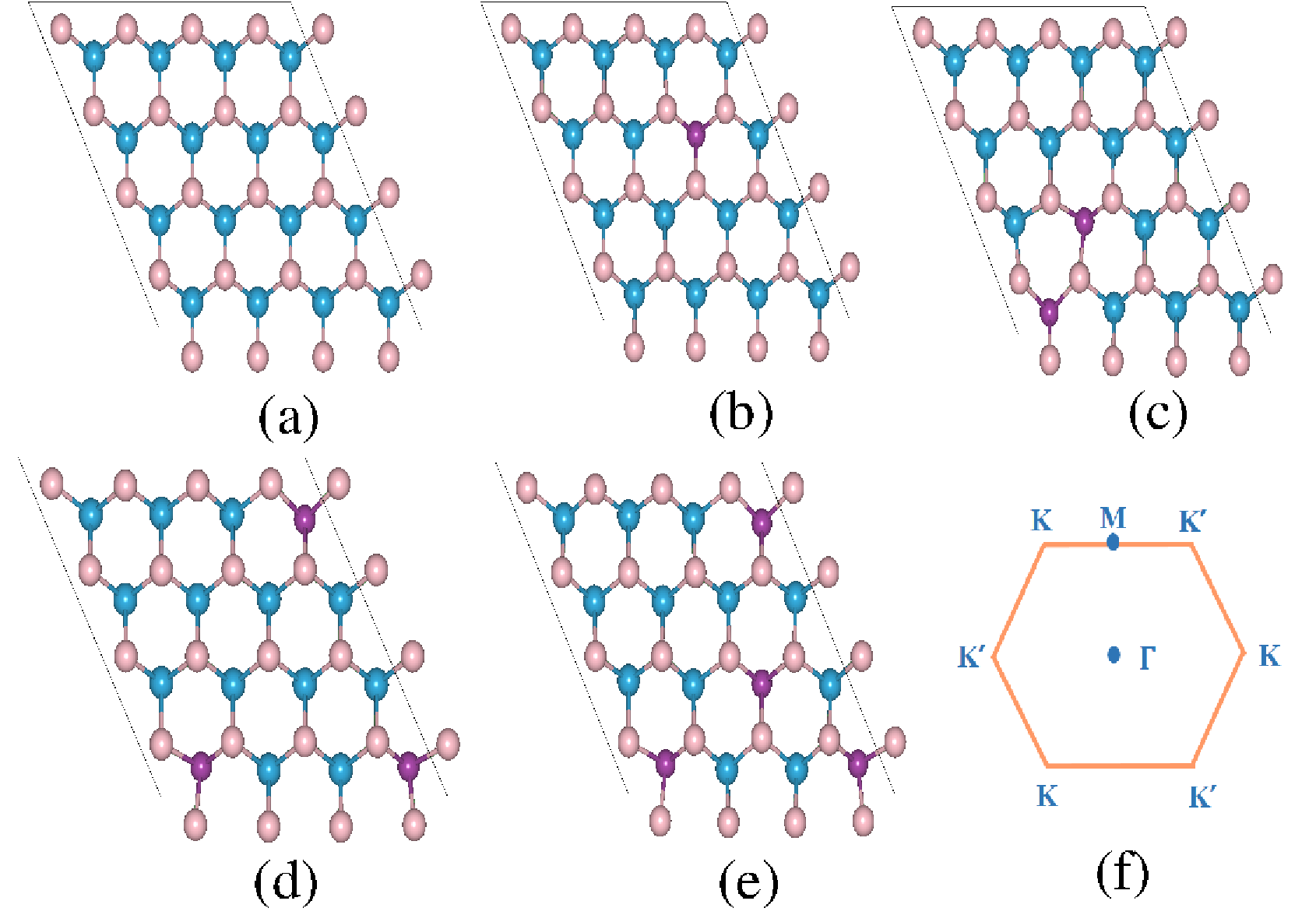}
	\caption{Top view of (a) crystal structure of pristine WSTe. panels (b), (c), (d),
	and (e) show the crystal structure of Fe-WSTe for 6.25, 12.5, 18.75,
	and 25\% concentrations. Blue, pink and purple spheres represent W, S, and Fe atoms, 
	respectively. (f) 2D Brillouin zone of WSTe monolayer showing 
	equivalent high-symmetry points.}
\label{fig_cryst}
\end{figure}

\section{RESULTS AND DISCUSSION}

\subsection{Crystal Structure and Structural Stability}

The Janus TMDs, represented as MXY (where M = Mo or W and X, Y = S, Se, or Te), 
exhibit a similar crystal structure as their parent material MX$_2$, in 
which metal atoms are placed between the two layers of chalcogen atoms and
form a hexagonal lattice. Therefore, the Janus monolayer structure originates 
from that of the TMDs with one layer of chalcogen atoms replaced with another 
group-VI elements. Bulk TMDs exist in 2H phase with space group D6h which 
possesses inversion symmetry. However, for a monolayer, the crystal symmetry 
is reduced to D3h for MX$_2$ and C3v for MXY, and the inversion symmetry is 
violated. The crystal structure of pristine WSTe Janus monolayer is 
shown in the panel (a) of the Fig. \ref{fig_cryst}.

We begin by optimizing the crystal structure of WSTe to achieve the global 
minimum ground state. For this, as there are no experimental inputs 
available for WSTe, we start with the experimental structure of
WS$_2$\cite{PhysRevB.7.3859} and perform full relaxation calculations after 
replacing one S atom with Te atom. The optimized structural parameters from 
our calculations are provided in Table \ref{tab_ene}.
Our computed lattice parameters for pristine WSTe are in good agreement 
with the previous calculations \cite{Er, PhysRevB.98.165424}.
From our calculations we find that, there is no significant change in the lattice 
parameters after the TM-substitutions. This is consistent with the 
previous theory calculation for TM-substituted WSSe \cite{ZHAO2019172}.
Table \ref{tab_ene} also displays the bond lengths and bond angles for TM-WSTe,
where L$_1$ and L$_2$ represent the bond lengths between W and S, and 
W and Te atoms, respectively. And $\theta$ is the angle between W and 
the surrounding S and Te atoms. As evident, both L$_1$ and L$_2$ decrease 
after the substitution of TM-elements, suggesting a stronger covalent 
interaction between the dopants and WSTe. A similar trend is also reported 
in TM substituted MoSeTe \cite{li2019spin}.
Since pristine WSTe has a graphene-like honeycomb structure, there are 
a series of equivalent high-symmetry $K$ points in the Brillouin zone, 
which are shown in panel (f) of Fig. \ref{fig_cryst} for the use in 
the later part of the paper for discussion. The panels (b), (c), (d), 
and (e) show the crystal structure of Fe-WSTe for 6.25, 12.5, 18.75,
and 25\% concentrations.

Further, to check the stability of TM-WSTe structures, we  examined the 
binding energy for all the considered concentrations. The data from 
this are given Table \ref{tab_ene}. The observed negative values of 
binding energy confirm the relative stability of TM-WSTe structures.

\begin{table*}
\caption{Calculated lattice parameters, Wyckoff positions, binding 
	energies, band gaps, spin polarization, and bond lengths 
	and bond angles for TM-WSTe structures.}
\centering
\begin{ruledtabular}
\begin{tabular}{llcccc}
	&   Lattice constants (\AA)  &  L$_1$ (\AA)  &  L$_2$ (\AA) & $\theta$ (degrees) \\
	WSTe    & 3.31 [Present work] & 2.429 & 2.718 & 83.11 & \\
	        & 3.35 \cite{Er} & & &   &\\
		& 3.36 \cite{PhysRevB.98.165424} & & &  & \\
	Fe-WSTe & 3.31 [Present work] & 2.282 & 2.623 & 82.44 & \\
	Co-WSTe & 3.31 [Present work] & 2.261 & 2.619 & 81.63 & \\
	Mn-WSTe & 3.31 [Present work] & 2.319 & 2.626 & 83.06 & \\
\hline
	Wyckoff positions &    & $x$  & $y$ & $z$ &    \\
	        & W  & 0.6667 & 0.3333&  0.7550 & \\
	        & S  & 0.3333 & 0.6667 & 0.8067 & \\
	        & Te & 0.3333 & 0.6667 & 0.6883 & \\ 
\hline 
        \% con.  & Space group     & Binding ene.      & Bandgap  & Phase & Spin-pol. \\
                 &      & (eV)    &  (eV)    &       & (\%) \\
\hline
 Fe-WSTe &&&&& \\
 6.25  & 156 & -37.220 & 0.000 ($\uparrow$), 1.150 ($\downarrow$) & Half-met.   & 100 \\
 12.50 & 8   & -40.984 & 0.051 ($\uparrow$), 0.536 ($\downarrow$) & Semic.   & - \\
 18.75 & 156 & -45.602 & 0.000 ($\uparrow$), 0.949 ($\downarrow$) & Half-met.   & 100 \\
 25    & 156 & -49.433 & 0.113 ($\uparrow$), 1.079 ($\downarrow$) & Semic.   & - \\ \\

 Mn-WSTe &&&&& \\
 6.25  & 156 & -39.329  & 0.102 ($\uparrow$), 1.208 ($\downarrow$)& Semic.   &  - \\
 12.50 & 8   & -44.249  & 0.024 ($\uparrow$), 0.593 ($\downarrow$)& Semic.   &  - \\
 18.75 & 156 & -50.938  & 0.024 ($\uparrow$), 1.140 ($\downarrow$)& Semic.   &  - \\
 25    & 156 & -56.315  & 0.000 ($\uparrow$), 1.161 ($\downarrow$)& Half-met.   & 100 \\ \\

 Co-WSTe &&&&& \\
 6.25  & 156 & -37.660 & 0.077 ($\uparrow$), 0.602 ($\downarrow$)  & Semic.    & -  \\
 12.50 & 8   & -39.782 & 0.051 ($\uparrow$), 0.067 ($\downarrow$)  & Semic.    & -  \\
18.75 & 156 &  -42.981 & 0.000 ($\uparrow$), 0.597 ($\downarrow$)  & Half-met. & 100 \\
25    & 156 &  -45.700 & 0.000 ($\uparrow$), 0.332 ($\downarrow$)  & Half-met. & 100 \\
\end{tabular}
\end{ruledtabular}
\label{tab_ene}
\end{table*}

\subsection{Electronic Structure and Rashba Spin Splitting}

\begin{figure}
\includegraphics[width=1.0
\columnwidth,angle=0,clip=true]{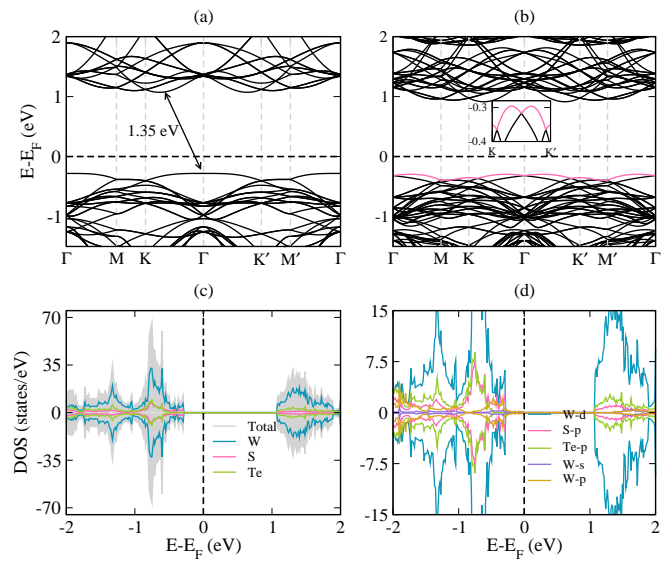}
\caption{The electronic band structure of WSTe, (a) without and (b) with 
	spin-orbit coupling. The inset in panel (b) shows the Rashba 
	splitting at the $\Gamma$ point of VBM. Panel (c) and (d) show the 
	atom and orbital-projected density of states, respectively. 
	The Fermi level is set to zero.} 
\label{fig_bands}
\end{figure}

Fig. \ref{fig_bands} shows the electronic band structure for pristine WSTe 
without and with spin-orbit coupling. As discernible from the panel (a) 
of the figure, WSTe is an indirect band gap semiconductor, with a computed 
band gap of 1.35 eV. 
The valence band maximum (VBM) lies at the $\Gamma$ point, 
whereas the conduction band minimum (CBM) lies between the $\Gamma$ and K points. 
However, when the spin-orbit coupling is switched on, the band gap decreases 
to 1.21 eV. Moreover, we observed distinct spin splittings at $K$ and 
$K^\prime$ valleys (panel (b)). We observed Rashba splitting 
at $\Gamma$ point, which will be discussed in detail in subsequent sections. 
The valence and conduction bands edges are observed to be dominated 
from 5d-orbitals of W, 3p-orbitals of S and 5p-orbitals of Te atoms.
The predicted electronic structure from our calculations for WSTe is in good 
agreement with the previous calculations \cite{Er, PhysRevB.98.165424}. 
To get further insight into the electronic structure, we examined 
the atom and orbital-projected density of states (pDOS) of WSTe. The 
data from this are shown in the panels (c) and (d) of the figure. The spin-up 
and spin-down states are observed to be symmetric, implying the non-magnetic 
nature of the pristine WSTe. Looking into the orbital project DOS, 
we find that, consistent with the band structure, for both valence and 
conduction bands, the dominant contribution comes from the $d$-electrons 
of W atoms. The next significant contribution is observed from the 
$p$-electrons of S and Te atoms.

In WSTe monolayer, the combination of time-reversal and broken inversion 
symmetries along with a strong spin-orbit coupling is reported to break 
the spin degeneracy in the valence and conduction bands. 
This can result into two types of spin splittings -- the Zeeman and Rashba spin splittings.
The Zeeman-type spin splitting occurs essentially due to the strong 
spin-orbit coupling arising from W atom and the absence of inversion symmetry \cite{PhysRevB.105.045426}. 
The Rashba-type spin splitting is, however, associated with the presence 
of an internal electric field perpendicular to the plane of the material \cite{PhysRevB.105.045426}. 
In the present case of WSTe, this internal electric is developed due to 
the difference in electronegativities of S and Te atoms.  
For a 2D material, the Rashba spin splitting can be explained using 
the Hamiltonian \cite{bychkov1984properties}

\begin{equation}
	H_{\rm R} = \alpha_{\rm R} ({\hat e}_z \times \vec k_{\parallel})\cdot\vec\sigma,
\end{equation}

where $\alpha_{\rm R}$ is referred to as the Rashba parameter and 
signifies the strength of the effect. And, ${\hat e}_z$, $\vec k_{\parallel}$, 
and $\vec\sigma$ represent the unit vector along field direction, momentum of 
electron, and the Pauli spin matrices, respectively.

\begin{figure}
\includegraphics[width=0.82
\columnwidth,angle=0,clip=true]{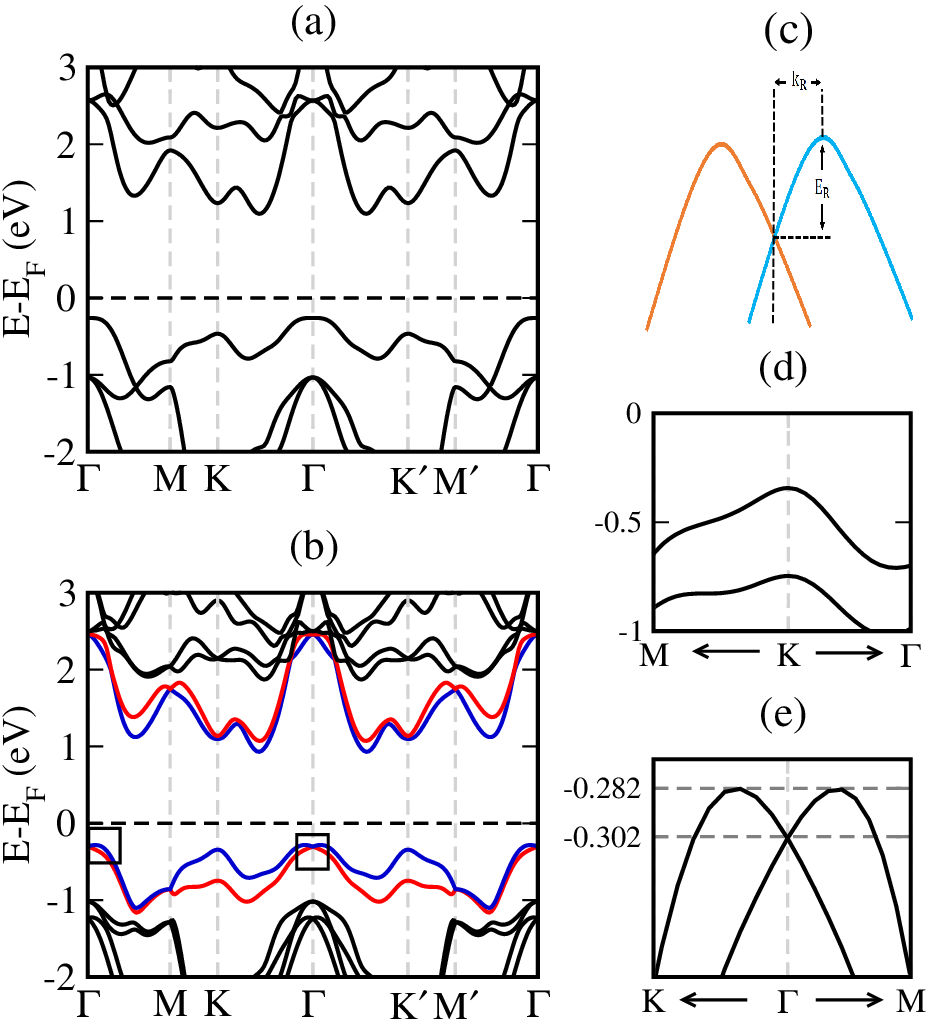}
\caption{The electronic band structure of WSTe unit cell, (a) without and 
	(b) with spin-orbit coupling. (c) A schematic digram showing Rashba 
	spin splitting.  (d) The zoomed view of Rashba splitting 
	at the $\Gamma$ point of VBM. (e) The zoomed view of Zeeman splitting 
	at $K$-point of VBM.}
\label{fig_rashba}
\end{figure}

\begin{figure}
\includegraphics[width=1
\columnwidth,angle=0,clip=true]{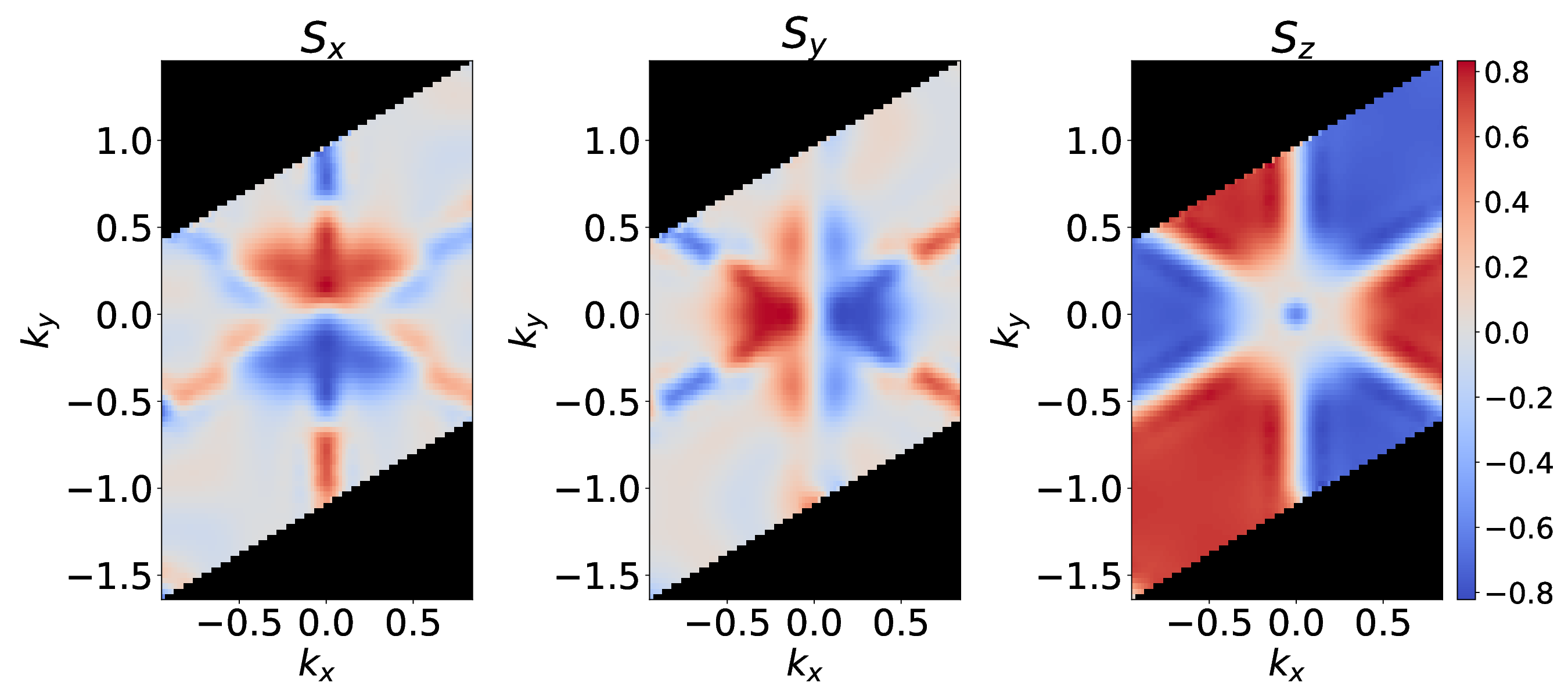}
\caption{The spin texture of VBM around $\Gamma$ point in $k_x - k_y$ 
	plane at 0.29 eV. Red and blue colors represent the 
	spin up and spin down states, respectively. }
\label{fig_spin}
\end{figure}

Fig. \ref{fig_rashba} shows the zoomed view of electronic band structure 
of WSTe unit cell without and with SOC. As can be observed from the panel (a) 
of the figure, as expected, all the bands are spin degenerate in the absence 
of SOC.  However, when the SOC is included, we observe a Rashba spin 
splitting around the $\Gamma$ and M points of the VBM and CBM, respectively. 
This is also confirmed from the spin texture plotted in $k_x - k_y$ 
plane around the $\Gamma$ point of the VBM (Fig. \ref{fig_spin}). 
As can be observed from the figure, near the zone center, the in-plane 
spin-polarized states dominate the texture, with almost no perpendicular 
component of the spin.
It should however be noted that the splitting around the M point of CBM 
is very small compared to that of around $\Gamma$ point in VBM. So, for 
further analysis and to quantify the effect we consider the splitting 
around $\Gamma$ point. 
The Rashba parameter could be obtained using the energy 
difference ($E_{\rm R}$) and the momentum offset ($k_{\rm R}$) using the 
relation $\alpha_{\rm R} = 2 E_{\rm R}/k_{\rm R}$ along both $\Gamma - K$
and  $\Gamma - M$ directions (Figs. \ref{fig_rashba}(b) 
and \ref{fig_rashba}(e)).
The values of $\alpha_{R}$ are obtained as 422 and 356 meV\AA\;
along $\Gamma - K$ and $\Gamma - M$ directions, respectively. These are 
in good agreement with the literature values of 322 and 324, 
respectively for WSTe \cite{PhysRevB.97.235404}. Our computed values 
of $\alpha_{\rm R}$ are also consistent with the reported values for 
other TMDs such as MoSeTe and WSeTe \cite{PhysRevB.97.235404}. It should be noted 
that a large Rashba spin splitting is preferable for spintronics based 
device applications. For Zeeman splitting, we observed a large spin splitting of ~ 403 meV 
at $K$ and $K^\prime$ valleys of the valence band (Fig. \ref{fig_rashba}(d)).
Compared to the valence band, the conduction band edges show a minor spin 
splitting of $\sim$37 meV. Our calculated values are in good agreement 
with the reported values, 426 and 29 meV, respectively, in the case 
of WSTe \cite{cheng2013spin}.

\begin{figure}
\includegraphics[width=0.9
\columnwidth,angle=0,clip=true]{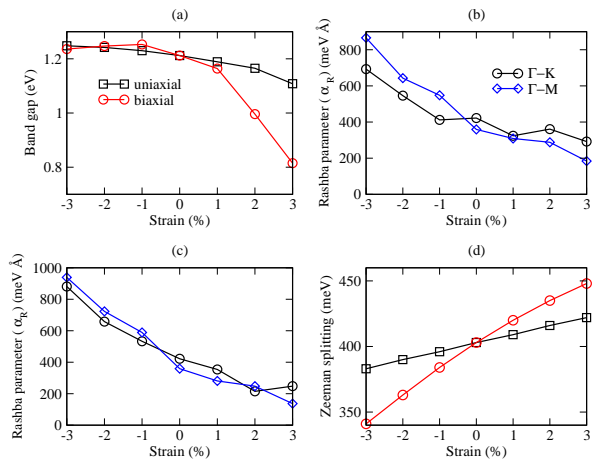}
	\caption{(a) The strain evolution of band gap.
	(b), (c) Rashba parameter under uniaxial and biaxial strains, respectively.  
	(d) Zeeman spin splitting in WSTe.}
\label{fig_strn}
\end{figure}

One of the efficient mechanisms to tune the magnitude of Rashba spin splitting 
in TMDs is reported through the application of 
strain \cite{PhysRevB.97.235404}.
The other proposed mechanisms could be the external electric field and 
charge dopings \cite{PhysRevB.87.035403,PhysRevLett.112.086802,D0RA00674B}.
Considering this, as the strain is reported to have more 
impact \cite{postorino2020strain,sun2019strain}, next we 
examine the effects of uniaxial and biaxial strains on Rashba and Zeeman-spin 
splittings. For this, we apply an uniaxial ($x$-direction) and biaxial compressive 
and tensile strains of up to 3\%. The strained evolved band gap,
Rashba parameter and Zeeman spin splitting energy from our simulations 
are shown in Fig. \ref{fig_strn}. As discernible from the panel (a) of the 
figure, band gap is observed to increase slowly with compressive strain, 
whereas the application of tensile strains reduces it, with biaxial 
strain showing more profound impact at larger strains.  
The observed trend is consistent with the findings in 
the literature \cite{el2023first}. The computed $\alpha_{\rm R}$ 
along both $\Gamma - K$ and $\Gamma - M$ directions are observed to show an overall 
increasing (decreasing) trend with compressive (tensile) uniaxial and 
biaxial strains (panels (b) and (c)). It is consistent with reported literature \cite{PhysRevB.97.235404}.
We find an enhancement of about 1.64\% and 2.43\% in 
$\alpha_{\rm R}$ for $\Gamma - K$ and $\Gamma - M$ directions, respectively for 3\% of 
compressive uniaxial strain. In the case of 3\% compressive biaxial strain,
an enhancement is about 2.08\% and 2.63\% for $\Gamma - K$ and $\Gamma - M$ directions, respectively.   
Interestingly, the Zeeman splitting show an opposite trend of evolution 
than Rashba splitting. We observe an almost a linear decrease (increase) in 
the splitting energy as a function of compressive (tensile) strains.

\begin{figure}
\includegraphics[width=1 \columnwidth,angle=0,clip=true]{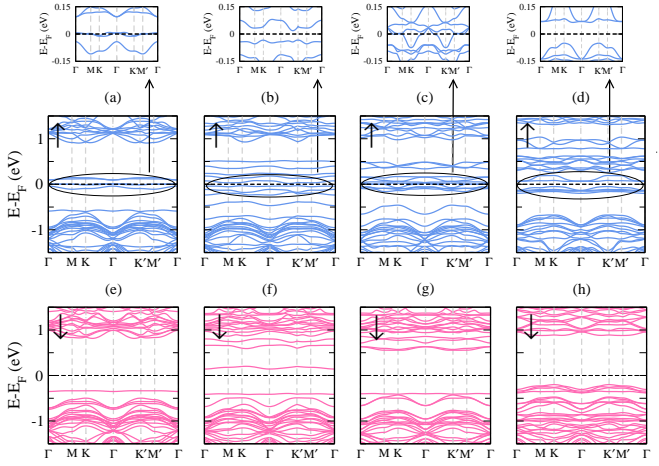}
\caption{The spin polarized band structure for Fe-WSTe without SOC, 
	(a) 6.25\%, (b) 12.5 \%, (c) 18.75\% , and (d) 25\% Fe-substitution, respectively.  
	The upper panels show the spin-up contribution, whereas the down
	panels show the spin-down states. The Fermi level is at zero.
	The zoomed view near Fermi level has also been shown.}
\label{fig_bands1}
\end{figure}

Next, we examine the electronic structure of TM-substituted WSTe. Fig. \ref{fig_bands1} 
shows the spin-polarized electronic band structure of Fe-WSTe for 6.25\%, 12.5\%,
18.75\%, and 25\% concentrations. The upper panels show the data for spin-up 
channel, whereas the lower panels represent the spin down channel.
As discernible from the figure, for 6.25 and 18.75\% concentration of Fe,
we observe a metallic behavior in the spin-up channel, whereas the spin-down 
channel exhibits a semiconducting behavior with band gaps of 1.15 and 0.95 eV
for 6.25\% and 18.75\%, respectively. This mixed nature of electronic 
structure suggests the half-metallic character of Fe-WSTe. Such materials 
could be crucial for promoting the spin-polarized transport in spintronics
applications \cite{choudhuri2019recent}.
On contrary, in the case of 12.5 and 25\% concentrations, both the
spin-up and down channels show a semiconducting nature; however, with a 
difference in the band gaps for spin-up and down channels. 
As evident from the Table \ref{tab_ene}, for 12.5\%, the spin-up channel
has a small band gap opening of $\sim$0.05 eV, whereas the spin down channel 
has an order more band gap of $\sim$0.54 eV.
Similarly, for the case of 25\% concentration, spin-up and down band gaps 
are $\sim$0.11 and $\sim$1.10, respectively. So, based on our predicted 
nature of electronic structure, one can conclude that the intrinsic 
properties of pristine WSTe could be tailored to exhibit either half-metallic 
or semiconducting characteristics depending on the need for device 
applications. The band gap modulation observed is significant for 
developing advanced spintronics devices which require a precise 
control over the spin-polarized transport \cite{liu2020spintronics}.

\begin{figure}
\includegraphics[width=1
\columnwidth,angle=0,clip=true]{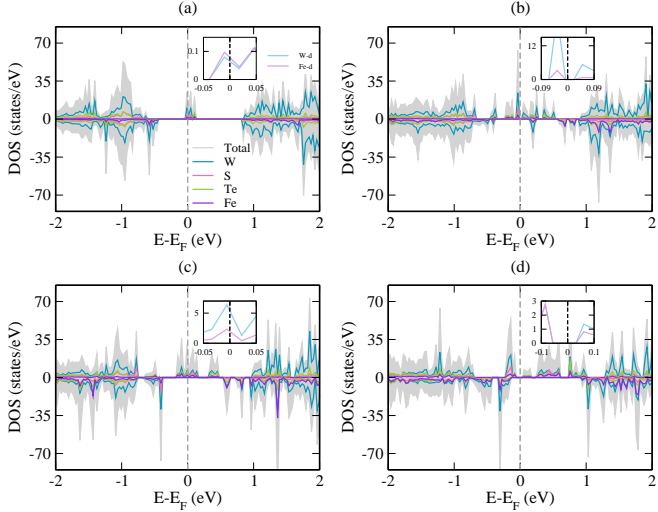}
\caption{The spin-polarized total and atom-projected density of states 
	for Fe-WSTe, (a) 6.25\%, (b) 12.5\%, (c) 18.75\%, and 
	(d) 25\%, respectively. The Fermi level is at zero.} 
\label{fig_pdos}
\end{figure}

To get insight into the origin of half-metallicity in Fe-WSTe, next we
examined the atom-projected density of states (pDOS) for Fe-WSTe. 
The data from our calculations are shown in Fig. \ref{fig_pdos} for 
all the concentrations. Comparing with pristine WSTe, we observed that 
the $5d$-electrons of W remain the dominant contributing states in both 
the valence and conduction bands for all the concentrations. Like W, Fe is 
also observed to contribute to both valence and conduction bands, however, 
with smaller magnitudes and more dominantly for the spin down channel.
Consistent with the band structure, for 6.25 and 18.72\% of Fe, for the 
majority spin, we observe non zero states at the Fermi level. As discernible 
from the insets of the figure, these contributions arise from the $3d$ and 
$5d$-electrons of Fe and W atoms, respectively. The absence of states at 
Fermi level for spin down channel leads to a 100\% spin polarization 
for these concentrations. For the 
case of 18.75 and 25\% concentrations, however, while spin down channel 
exhibits no states at Fermi level, there is a small band gap opening 
in the majority spin. Our obtained electronic structure of Fe-WSTe is 
consistent with the reported data for another material of Janus family, 
Fe-MoSeTe \cite{li2019spin}

\subsection{Magnetic properties}

Next, we present and analyze our data on magnetic properties 
of TM-WSTe. The TM substitutions in non-magnetic 2D 
materials are observed to introduce magnetic degree of freedom 
in the materials \cite{anagha2024effect,fu2020enabling}.
First we investigate the ground state magnetic configuration of TM-WSTe systems. 
For this, we examined both ferromagnetic (FM) and antiferromagnetic (AFM) 
orientations of magnetic moments for all the Fe/Mn/Co concentrations. 
In Table \ref{tab_mag}, we have provided the relative energy for 
FM phase with respect to AFM. As evident from the table, our calculations 
reveal the actual ground state of Fe-WSTe as a FM phase for all the concentrations. 
The FM energy is found to be $\approx$ 0.55, 0.28, 0.12, and 0.37 eV lower than that 
of AFM phase for 6.25, 12.5, 18.75, and 25\% concentrations, respectively.
We observe a similar trend for Mn-WSTe and Co-WSTe systems, where FM 
configuration is observed to be the actual ground state magnetic 
configuration, except for 25\% Co-WSTe.

\begin{figure}
\includegraphics[width=0.85 \columnwidth,angle=0,clip=true]{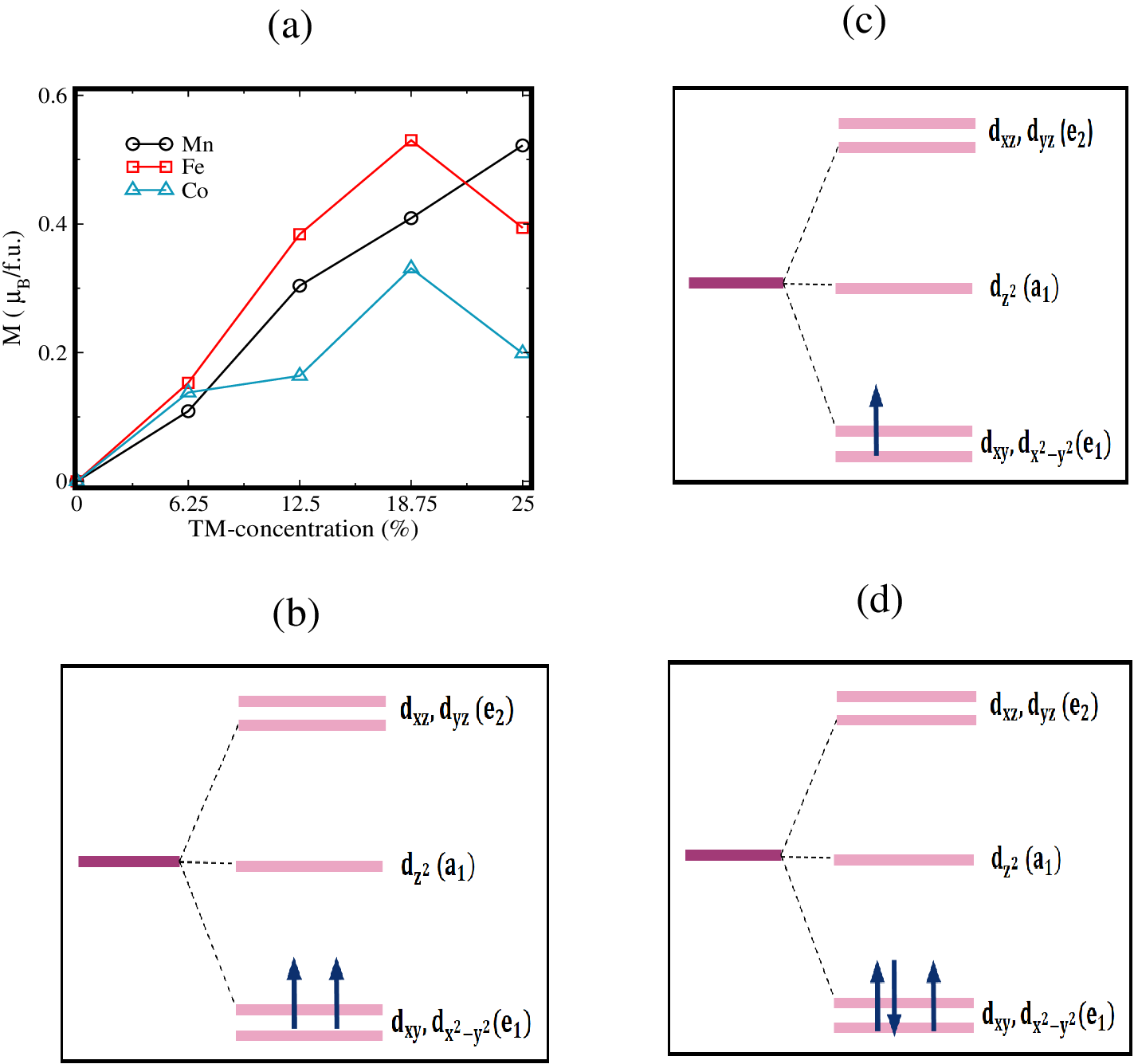}
	\caption{(a) Magnetic moment as a function of Fe, Mn and Co concentration
	in TM-WSTe. Octahedral filling of $d$-electrons in Fe-WSTe (panel (b)), 
	Mn-WSTe (panel (c)), and Co-WSTe (panel (d)) systems.}
\label{fig_mag}
\end{figure}

Examining the total magnetic moments provided in Table \ref{tab_mag}, as can 
be expected, our simulations predict a zero magnetic moment for pristine WSTe, 
which is in agreement with the literature data \cite{absor2018tunable,tang20222d}. 
However, for TM-WSTe, we observe nonzero magnetic moments with a trend 
of increasing magnitudes, except for 25\% in the case of Fe and Co, 
with concentrations (panel (a) of Fig. \ref{fig_mag}). 
The reason for the increase in total magnetic moment
could be attributed to the increasing ferromagnetic exchange between 
neighboring Fe/Mn/Co ions at higher concentrations. The decrease in 
the value of total magnetic moment in Fe and Co for 25\% concentration
could be attributed to the reduction in the magnetic moments of 
Fe and Co atoms due to an increased hybridization between Fe/Co-$3d$ 
and WSTe states (specially $p$-electrons of S and Te atoms).
Also, at this concentration, W, S, and Te atoms have more 
opposite contributions, which reduces the total magnetic moment.

For Mn-WSTe, the largest total magnetic moment is obtained as 0.522 $\mu_B$/f.u. 
for the highest concentration of 25\%.
Similarly, for Fe-WSTe and Co-WSTe, these are 0.530 and 0.331 $\mu_B$/f.u.,
respectively, for 18.75\% concentration. 
Our obtained magnetic moments for Fe/Mn/Co-WSTe for 6.25\% are
in good agreement with the previous study on parent compound 
TM-WS$_2$ \cite{yang2016effect}.
As can be observed from the 
table, the dominant contribution to total magnetic moment comes from 
the TM-elements. 
For example, for 6.25\%, we observe $\approx$ 133, 189, and 66\% 
of contributions to total magnetic moment from Fe, Mn, and Co, respectively.
The reason for this dominant contribution could be ascribed to the presence 
of unpaired $d$-electrons in these elements. This is shown in the 
octahedral filling of $3d$-electrons in Fe/Mn/Co-WSTe in Fig. \ref{fig_mag}. 
Interestingly, there are significant contributions of opposite phase 
from W, S and Te atoms through the 
interaction with TM-elements. The opposite sign leads to cancellations, 
and hence reduces the total magnetic moment. 

\begin{table}
        \caption{The relative energy ($\Delta E = E_{\rm FM} - E_{\rm AFM}$)
        (in eV) of FM and AFM configurations and the atom resolved magnetic
        moments (in $\mu_{\rm B}/atom$) for TM-substituted WSTe for all
        considered concentrations.}
\centering
\begin{ruledtabular}
\begin{tabular}{lrrrrcr}
\% con. &  $\Delta E$ & $\mu^{\rm W}$ & $\mu^{\rm S}$ & $\mu^{\rm
        Te}$ & $\mu^{\rm Fe/Mn/Co}$ & $\mu^{\rm Tot}$ \\ \hline
WSTe  &  & 0.0 & 0.0  & 0.0 & 0.0 & 0.0 \\

Fe-WSTe &&&&&& \\
6.25  & -0.559 & -0.047 & -0.000 & -0.007 & 3.276 & 2.451 \\
12.5  & -0.285 & -0.014 & -0.003 & -0.000 & 3.202 & 6.139 \\
18.75 & -0.129 & -0.041 & -0.014 & -0.007 & 3.127 & 8.482 \\
25    & -0.376 & -0.150 & -0.061 & -0.079 & 2.590 & 6.310 \\

Mn-WSTe &&&&&& \\
6.25  & -1.063 & -0.050 & -0.021 & -0.028 & 3.308 & 1.749 \\
12.5  & -0.244 & -0.058 & -0.035 & -0.040 & 3.445 & 4.864 \\
18.75 & -0.192 & -0.136 & -0.058 & -0.063 & 3.429 & 6.559 \\
25    & -0.264 & -0.174 & -0.077 & -0.094 & 3.392 & 8.358 \\

Co-WSTe &&& &&& \\
6.25  & -0.125 &  0.063   & -0.014  &  0.002  &  1.454  & 2.212   \\
12.5  & -0.092 & -0.014   & -0.025  & -0.013  &  1.732  & 2.636   \\
18.75 & -0.026 & -0.028   & -0.008  &  0.001  &  1.925  & 5.296   \\
25    & +0.017 & -0.051   & -0.066  & -0.035  &  1.358  & 3.187    \\
\end{tabular}
\end{ruledtabular}
\label{tab_mag}
\end{table}

\begin{figure}
\includegraphics[width=1.0 \columnwidth,angle=0,clip=true]{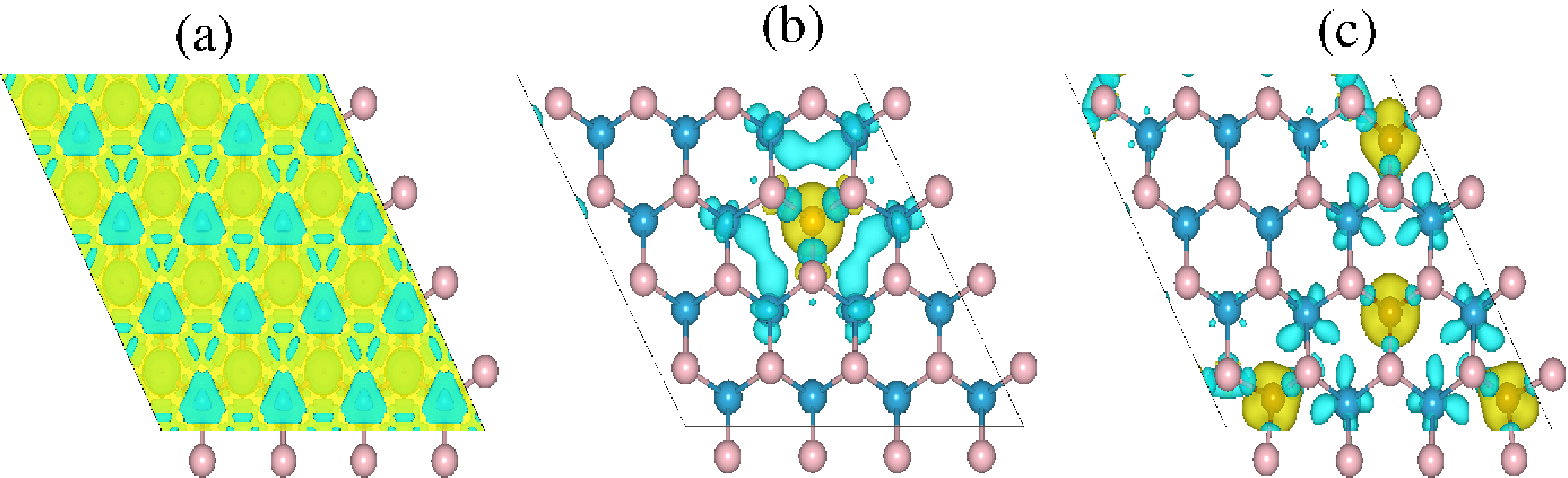}
\caption{Isosurface plots for the spin charge density, (a) pristine WSTe 
	(b) 6.25\%, and (d) 25\% Fe-WSTe monolayer. The Isosurface value
	is extracted to be 0.003 $e/\text{Å}^3$.} 
\label{fig_charg}
\end{figure}

To get further insight into the spin orientation and resultant nonzero 
magnetic moments in TM-WSTe, next we examined the spin density distribution in
these systems. Fig. \ref{fig_charg} shows the Isosurface plots for the spin 
charge density of Fe-WSTe for pristine WSTe, and 6.25 and 25\% concentrations. 
The spin charge density is defined as $\Delta\rho = \rho_{\uparrow} - \rho_{\downarrow}$, 
where $\rho_{\uparrow}$ and $\rho_{\downarrow}$ represent the spin-up and 
spin-down charge densities \cite{ZHAO2019172}. 
As can be expected, for pristine WSTe, the spin density distribution is 
uniform across the plane of the material (panel (a)), and as a result it 
leads to the zero spin polarization. For Fe-WSTe, however, the spin density 
distribution is observed to be more localized with Fe atoms. This suggests 
a charge transfer between TM elements and WSTe, leading to a nonzero spin
polarization in  the system.

\subsection{Valley Polarization}

\begin{figure}
\includegraphics[width=1
\columnwidth,angle=0,clip=true]{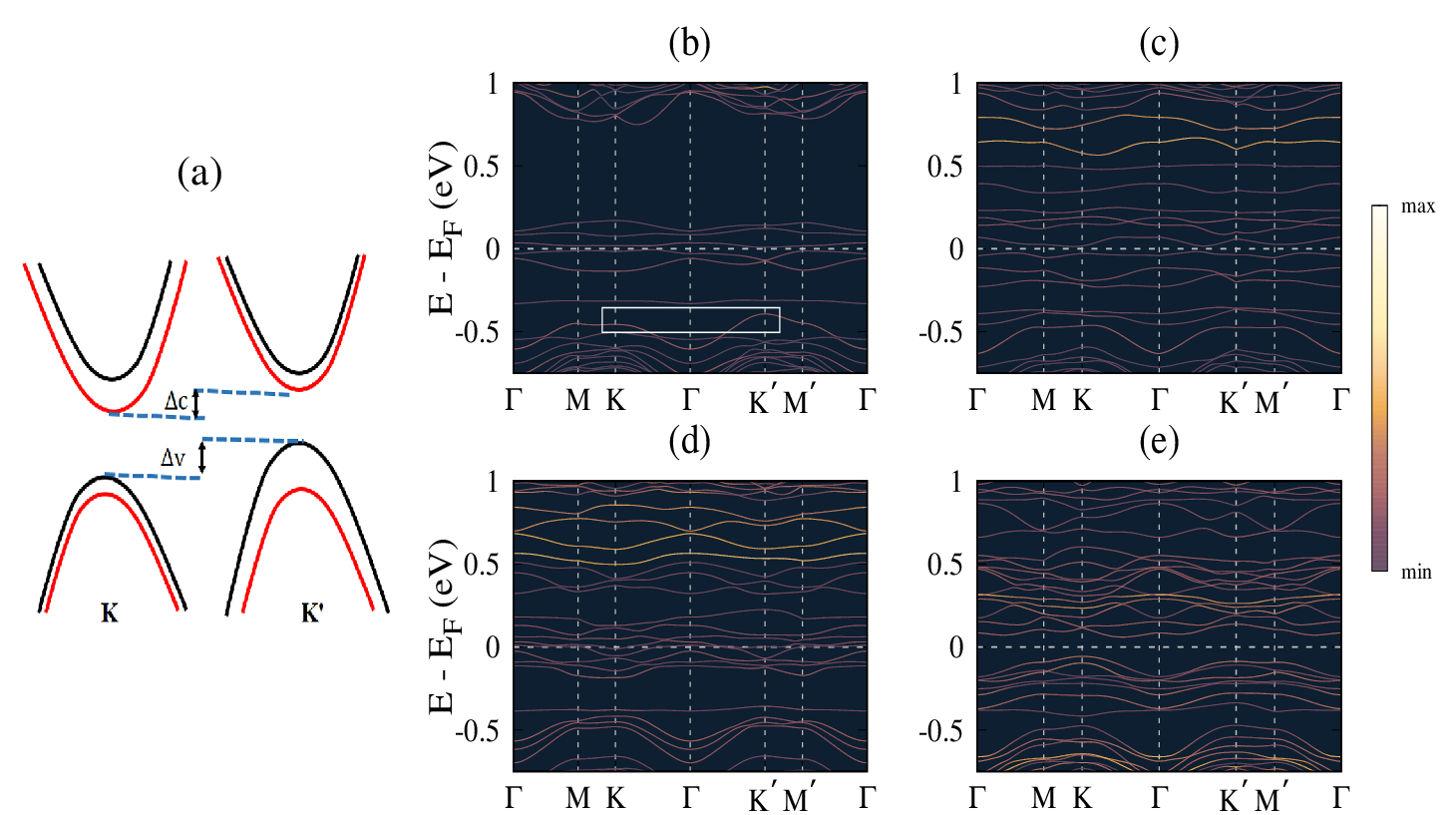}
\caption{(a) A schematic diagram representing the valley 
	polarization in TM-WSTe.  
	The spin-orbit coupled band structures of Fe-WSTe for 
	6.25 (panel (b)), 12.5 (panel (c)), 18.75 (panel (d)), 
	and 25\% (panel (e)) concentrations. The rectangular 
	box is to indicate the band used in the calculation of 
	valley polarization.}
\label{fig_valley}
\end{figure}

Next, we present and discuss our simulation results on the valley 
polarization. Valley polarization in 2D materials is induced as 
a combined effect of broken time reversal symmetry, which could be
introduced via TMs dopings, and a strong spin-orbit coupling \cite{ZHAO2019172}.
The schematic representation of valley polarization is shown in panel (a)
of Fig. \ref{fig_valley}. The valley polarization can be quantified in terms 
of the energy difference, $\Delta_{KK'} = \left| E_{K'} - E_{K} \right|$, 
between the $K$ and $K^\prime$ valleys. Panels (b, c, d, e) of Fig. \ref{fig_valley} 
shows the electronic band structure of 
Fe-WSTe in the presence of SOC for all the chosen concentrations. We have two key 
observations from the figure. First, the defect states 
emerge in the vicinity of the Fermi level. These defect states
mainly comprise of W and Fe atoms, which suggests a strong orbital 
hybridization between these atoms. And second, the spin degeneracy in 
the bands is lifted. Because of the opposite spins associated with $K$ 
and $K^{'}$ valleys due to broken time-reversal symmetry, there is an 
uneven splitting of the energy levels at these valleys of VBM and CBM. 
As can be observed from panel (b), for 6.25\% Fe-WSTe, the spin splitting 
at CBM is much smaller than that at the VBM. The reason for this 
could be attributed to the different orbital's contribution to the 
edges of VBM and CBM -- $d_{z^2}$ is observed to have dominant 
contribution to CBM, whereas the VBM is populated mainly by $d_{xy}$ 
and $d_{x^2 - y^2}$ orbitals. 
As can be observed from panel (b), for 6.25\% concentration, the spin 
splitting at $K$ valley is larger than that at $K^\prime$, 
leading to a valley polarization of $\sim$ 65 meV. This observed 
value of valley polarization from our calculations for Fe-WSTe is 
significantly higher than the previously reported values, 58 meV, for 
transition metal (V) doped-WSSe \cite{ZHAO2019172}.
For higher concentrations (panels (c), (d), and (e)), we observe 
an increase population of defect states in the vicinity of Fermi level 
and within the band gap. This leads the deformation of valleys and makes 
it difficult to distinguish between the defects and valley-polarized 
states. So, as a result, it is extremely challenging to accurately 
quantify the valley polarization for these concentrations.

\begin{figure}
\includegraphics[width=1
\columnwidth,angle=0,clip=true]{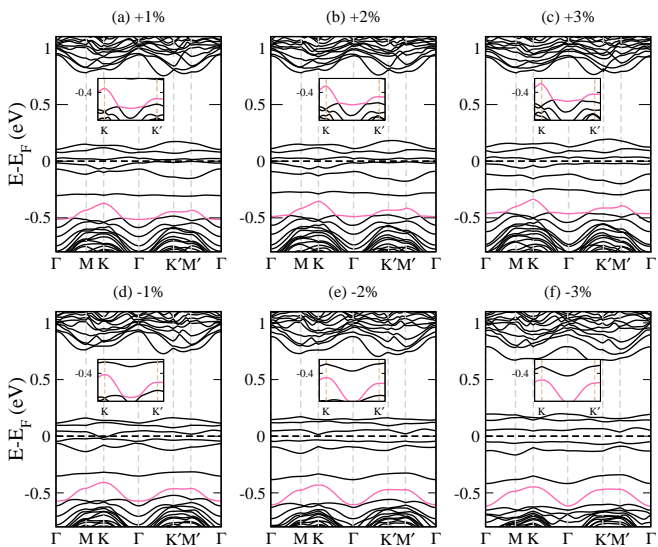}
\caption{Electronic band structure of 6.25\% Fe-WSTe under the uniaxial 
	strain, (a) +1\% (b) +2\% (c) +3\% (d) -1\% (e) -2\% and (f) -3\%. 
	The Fermi level is at 0 eV.} 
\label{fig_uni}
\end{figure}

\begin{figure}
\includegraphics[width=0.75
\columnwidth,angle=0,clip=true]{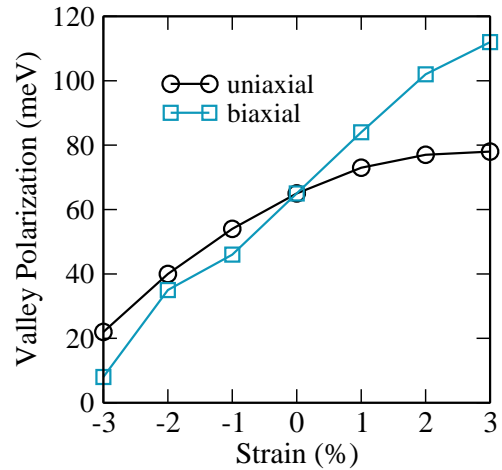}
\caption{Valley polarization as a function of uniaxial and biaxial strain 
	 for 6.25\% concentration in Fe-WSTe.}
\label{fig_valley2}
\end{figure}

Since the valley polarization is a key parameter and governs various 
phenomena like anomalous valley Hall effects, valley transport, 
valley-polarized superconductivity and valley-based quantum 
computing \cite{tong2016concepts,vitale2018valleytronics}, it is crucial 
to explore the mechanisms to enhance it. One of the most common and 
feasible approach to enhance valley polarization effects in 2D materials 
is through the application of strain \cite{hsu2020nanoscale,zheng2023strain}. 
Considering this, next we examine the effects of uniaxial and biaxial 
strains on valley polarization properties of TM-WSTe. 
Fig. \ref{fig_uni} shows the electronic band 
structure for 6.25\% Fe-WSTe at different compressive and tensile 
uniaxial strains. As can be observed from the figures, the energy 
difference between the $K$ and $K^\prime$ valleys at the VBM 
increases (decreases) with increasing tensile (compressive) strain. 
This leads to an increase (decrease) in the valley polarization values 
as function of tensile (compressive) strain (Fig. \ref{fig_valley2}(a)). 
The maximum valley polarization is obtained as 78 meV for 3\% of tensile 
strain. We observe a similar trend for $\Delta E_{KK^\prime}$ as the 
function of biaxial strain also (Fig. \ref{fig_valley2}(a)), however, 
with a key difference. The change in the valley polarization with 
strain is more profound in the case of biaxial strain. We could achieve 
a maximum valley polarization of 112 meV with an application of 
3\% biaxial tensile strain. These findings suggest that with the 
help of moderate strain engineering 
one can tune the valley polarization of TM-WSTe. Our results further 
highlight that Fe-WSTe could be a promising candidate for 
valleytronic applications.
In contrast to traditional charge-based electronics, valleytronics 
offer a low-energy information processing, as the transition between 
valley states requires minimal 
energy \cite{renard2015valley,sakamoto2013valley,luo2024valleytronics}. 
Real-time control of valley polarization can be used in high-speed transistors, 
non-volatile memory devices, and innovative quantum computing applications 
\cite{rohling2012universal,rycerz2007valley,gunawan2006quantized}.

\begin{table}[ht]
	\caption{Valley polarization at different uniaxial and biaxial 
	strains for 6.25\% Fe-WSTe.}
\centering
\begin{ruledtabular}
\begin{tabular}{cccc}
	Uniaxial Strain (\%) & $\Delta_{KK'}$ (meV) & Biaxial Strain (\%) 
	& $\Delta_{KK'}$ (meV) \\
\hline
-3 & 22 &  -3 & 8 \\
-2 & 40 &  -2 & 35 \\
-1 & 54 &  -1 & 46 \\
0  & 65 &   0 & 65 \\
+1 & 73 &  +1 & 84 \\
+2 & 77 &  +2 & 102 \\
+3 & 78 &  +3 & 112 \\
\end{tabular}
\end{ruledtabular}
\label{tab_vp}
\end{table}

\begin{figure}
\includegraphics[width=1
\columnwidth,angle=0,clip=true]{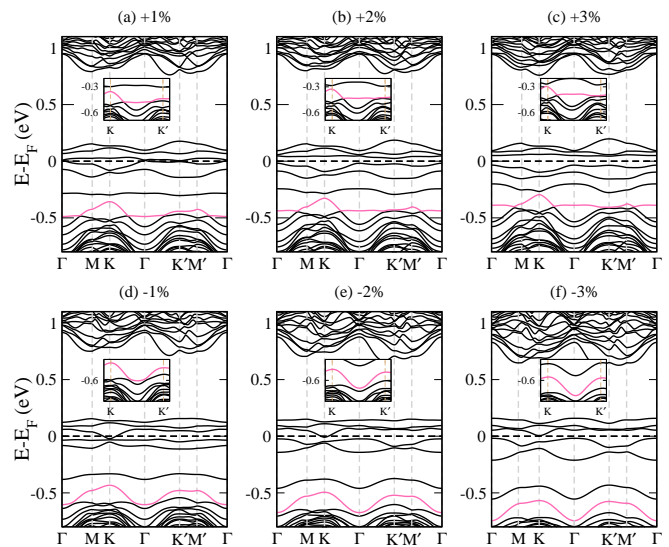}
\caption{Electronic band structure of 6.25\% Fe-WSTe under the biaxial 
	strain, (a) +1\% (b) +2\% (c) +3\% (d) -1\% (e) -2\% and (f) -3\%. 
	The Fermi level is at 0 eV.} 
\label{fig_bi}
\end{figure}

\section{CONCLUSIONS}

In summary, with the help of density functional theory based first-principles 
calculations, we have examined the effect of transition metal substitution 
on the electronic, magnetic, and valleytronic properties of WSTe monolayer.
In agreement with the literature data \cite{PhysRevB.97.235404}, our 
simulations on electronic structure predict WSTe as an indirect bandgap 
semiconductor. Our computed bandgap, 1.35 eV, agrees well with previous 
data \cite{PhysRevB.97.235404}. The WSTe monolayer is found to inherently 
exhibits Rashba and Zeeman-spin splittings due to the intrinsic out-of-plane 
electric field and in-plane inversion symmetry breaking, respectively, 
along with a strong spin-orbit coupling. Our obtained Rashba parameter and 
Zeeman-spin splitting are 422 meV\AA\; and 403 meV, respectively. 
Our strain dependent calculations show an increasing(decreasing) trend of
Rashba parameter with compressive(tensile) uniaxial and biaxial strains. 
Interestingly, the Zeeman splitting is observed to show a linear 
dependence on strain with an opposite trend of decreasing(increasing) energy with
compressive(tensile) strains.

From the electronic structure of TM-WSTe, we observed an emergence of half-metallic 
ferromagnetism with 100\% spin polarization for 6.25 and 18.75\% of Fe, 
25\% of Mn, and 18.75 and 25\% of Co structures. The largest effective 
magnetic moments for supercell are obtained as 8.5, 8.4 and 5.3 
$\mu_{\rm B}$ for Fe/Mn/Co-WSTe structures, respectively. 
Our simulations on valley polarization predict the values 65, 54.4 and 46.3 
meV for 6.25\% of Fe, Mn and Co substitution, respectively. Our 
calculations show a strain dependent tunability of valley polarization, 
where it is observed to increase(decrease) with tensile(compressive) uniaxial 
and biaxial strains. For Fe-WSTe, a maximum valley polarization of 112 is
obtained with an application 3\% biaxial strain.
The observed significant values of spin-polarization, Rashba splitting 
and valley polarization suggest that TM-WSTe could offer a potential
candidate for spintronics and valleytronics applications.

\section*{ACKNOWLEDGMENTS}

BKM acknowledges the funding support from SERB, DST (CRG/2022/000178).
Shivani acknowledges the fellowship support from UGC (BININ01949131), 
Govt. of India.
The calculations are performed using the High Performance Computing cluster Tejas
at the Indian Institute of Technology Delhi and PARAM Rudra, IUAC, New Delhi.

\bibliography{WSTe}

\end{document}